# Influence of Roughness on Granular Avalanches

Jie Zheng[1], Yi Xing[1], Ye Yuan[1], Zhifeng Li[1], Zhikun Zeng[1], Shuyang Zhang[1], Houfei Yuan[1], Hua Tong[1], Chengjie Xia[2*], Walter Kob[1,3], Xiaoping Jia[4], Jie Zhang[1,5] and Yujie Wang[1,6*]

[1]*School of Physics and Astronomy, Shanghai Jiao Tong University, Shanghai 200240, China*
[2]*School of Physics, East China Normal University, Shanghai 200241, China*
[3]*Laboratoire Charles Coulomb, University of Montpellier and CNRS, Montpellier 34095, France*
[4]*Institut Langevin, ESPCI Paris, Université PSL, CNRS, Paris 75005, France*
[5]*Institute of Natural Sciences, Shanghai Jiao Tong University, Shanghai 200240, China*
[6]*Materials Genome Initiative Center, Shanghai Jiao Tong University, Shanghai 200240, China*

Combining X-ray tomography with simultaneous shear force measurement, we investigate shear-induced granular avalanches using spherical particles with different surface roughness. We find that systems consisting of particles with large surface roughness display quasi-periodic avalanches interrupted by crackling-like small ones. In contrast, systems consisting of particles with small roughness display no detectable avalanches. The stress drop of quasi-periodic avalanche shows a linear relation with the correlation length of particle non-affine displacement, suggesting that roughness enhances inter-particle locking and hence particle-level dynamic correlation length. However, the nonaffine displacement is two orders of magnitude smaller than particle size, indicating that stress is mainly released on the length scale of roughness. The correlation length of non-affine displacements abruptly increases when a quasi-periodic avalanche occurs, suggesting that quasi-periodic avalanches can be interpreted as a spinodal nucleation event in a first-order phase transition.

Granular materials often display significant mechanical and structural fluctuations during shear deformation or flow [1-4]. These fluctuations are intermittent and consist of characteristics of avalanche behavior such as abrupt surface slips or structural rearrangements [3-5]. Granular avalanches are ubiquitous in nature and related to many geological catastrophic processes including earthquake, landslide, etc. [6-8]. It is found that friction [9], particle roughness [10, 11], shear rate [12], loading stiffness [12, 13], boundary roughness [14], particle shape [15, 16] and inertia effect [12], etc., can influence granular avalanches in complex ways, making that the essential physics that governs granular avalanches has remained elusive [3, 5, 17].

The global stress (or shear force) drop $s$ is one of the most important variables to characterize avalanches [5]. Pervious experimental studies mainly focused on the power law distribution of $s$ [13, 16, 18-20], *i.e.*, $p(s) \sim s^{-1.5} \exp(-s/s_c)$, with few studies addressing the connection between particle-scale microscopic dynamics and global stress fluctuations for three dimensional (3D) granular systems [16, 19]. Additionally, different from simple scale-free or truncated power-law distributions, granular matter and earthquake faults also often display a marked hump tail on their $p(s)$ curves during avalanche [21-24]. This feature indicates a characteristic, quasi-periodic avalanche behavior, which so far has received little experimental attention [21, 22, 25]. Previously, dynamic weakening and the narrow/wide distribution of system heterogeneity have been proposed to rationalize the occurrences of these quasi-periodic avalanches and their corresponding phase behaviors [6, 26, 27]. Yet, these theoretical ideas have so far not been tested in experiments. In particular, the microscopic dynamics that governs quasi-periodic granular avalanche, the relevant length scale on which it occurs, and the related pathway leading to an avalanche remains under intense debate [23, 26, 28, 29].

Avalanches in granular systems belong to the general family of avalanche behaviors which covers a wide field including the fracture of materials, the yielding of fault zones, Barkhausen noise in magnets, the plasticity of amorphous solids, etc. [5, 6, 30, 31]. Mean-field theories and numerical studies suggest that, essentially, an avalanche behaves like a phase transition process, in which the avalanche size distribution and the nature of the transition (first-order or second-order) are governed by the competition between interaction strength and inherent disorder [32-35]. Here, the interaction strength could be controlled by interaction potential, frictional resistance, etc., and the system disorder by structure or stress heterogeneity, or local yield stress distribution. Specifically, when interaction dominates disorder, shear localization will occur [33, 36], and the avalanche occurs through the nucleation of dynamic clusters, analogous to a first-order phase transition [28, 29, 33, 37]; inversely, the avalanche will show a second-order phase transition behavior, *e.g.*, a percolation transition, with spatially more sporadic dynamic clusters and a power law divergence of the mean cluster size as an avalanche is approached [19, 33, 38, 39]. However, in reality the system often displays mixed types of phase behaviors [26, 27, 33], and only under special conditions, the system will display a scale-free distribution of stress drops, reminiscent of behaviors of self-organized criticality [32, 33]. It is therefore crucial to validate these theoretical concepts by experimental studies on granular avalanches, and also to explain how specific granular properties influence the avalanche type.

In this work, avalanches in 3D granular systems which are slowly sheared (the corresponding inertial number is $I = \dot{\gamma}d/\sqrt{P/\rho} \approx 2.4 \times 10^{-7}$) with uniform strain steps are comprehensively studied by X-ray tomography (CT) technique and simultaneous shear force measurements. Three different surface roughness particles have been investigated. We find that systems with large roughness (*LR*) particles display quasi-periodic stress drops (avalanches) interrupted by smaller, intermittent ones. The systems with medium roughness (*MR*) particles display weakened and more frequent quasi-periodic avalanches, while the systems with small roughness (*SR*) display no detectable avalanche within experimental resolution. We find that the quasi-periodic avalanche strength is directly proportional to the correlation length $\xi$ of particle-level nonaffine dynamics. In addition, we also find that as an avalanche is approached, $\xi$ first increases slowly, and then abruptly rises to a large value at the critical strain when avalanche occurs. This suggests a spinodal-like nucleation picture for quasi-periodic avalanche which is of first-order phase transition nature.

The shear cell has a 3D simple shear geometry, as shown in Fig. 1. A step motor attached to the bottom plate of the shear cell through a linear stage is used to shear the system. A force sensor (eDPU-200N, IMADA Company) connected between the motor and the stage is used to measure the shear force. The sensitivity of the force sensor is 0.1N, and the maximum sampling rate is 10kHz. The base frictional force of the shear cell with no particles inside is about 1.5N. A heavy block is placed on the top of the shear cell through ball bearing channels to impose a constant pressure about 0.7kPa on the packing, which can freely move in both vertical and horizontal directions to accommodate shear deformation of the system.

The sample contains about 13,000 bi-dispersed hard spherical spheres, with equal number of 5mm and 6mm diameter particles. The mean particle diameter is set as a unit length *D*. All the particles are 3D printed (ProJet 2500Plus) using plastic materials, with stiffness 1.75GPa and friction coefficient about 0.6. *LR* and *MR* systems use 3D-printed particles from different batches and particles in *SR* are obtained by machine polishing of *LR*. The insets of Fig. 2(a) show optical and scanning electronic microscope (SEM) images of these particles, displaying distinctive surface morphologies. The roughness is estimated by surface unevenness from SEM images: ~50μm for *LR*, ~30μm for *MR* and <10μm for *SR* systems. This experimental approach enables us to isolate the influence of particle roughness on granular avalanche by eliminating other potential contributing factors such as particle shape [15, 16], shear rate [12, 13], loading stiffness [13], etc.

Before each experiment, a reproducible initial state is prepared by applying 300 cycles of cyclic shear with amplitude $\gamma = \pm 10\%$ on the packing. Subsequently, simple shear with uniform strain steps ($\delta\gamma = 0.91\%$) is applied to the system until a maximum strain of 50%. After each step strain, the microscopic packing structure is acquired by X-ray tomography (UEG Medical Group Ltd., 0.2 mm spatial resolution), and then analyzed by an image processing algorithm following previous studies [40, 41]. For each type of system, we have eight simple shear realizations for better statistics.

We first show global stress-strain curves for three systems in Fig. 2(a). Interestingly, the properties of the avalanches strongly depend on the roughness of the particles: for *LR* and *MR* systems, the avalanches are quasi-periodic, interrupted by weak stress drops; while for *SR* system, the avalanches are much weaker or cannot be resolved experimentally. Figure 2(a) also shows the evolutions of corresponding global volume fractions $\phi$. All of them approach the steady state with $\phi \approx 0.6$ in a smooth manner unlike the jerky behavior of the shear forces. In Fig. 2(b), the top panel shows the probability distribution function (pdf) of stress drop *s* for the *MR* system, which displays a power law truncated by a hump tail. The smaller avalanches (roughly, $s < 10\text{N}$) satisfy a power law distribution of $p(s) \sim s^{-1.5}$, which resembles scale-free crackling noise [13, 30]. Crackling-like avalanches have been well documented and explained by both experiments [16, 18, 19] and mean-field theories [5, 6, 17, 30, 37]. However, unlike usual truncated power law distribution [16, 17, 19], the appearance of a hump tail indicates the existence of a quasi-periodic avalanche with a characteristic stress drop magnitude [21, 22, 24, 27]. For the avalanche events in the hump tail of $p(s)$ curve, we calculate the strain intervals between two successive drops (see the shaded area in top panel of Fig. 2(a) for the definition of strain interval for a typical avalanche event). The histograms of these strain intervals display characteristic peaks for both *LR* and *MR* systems (middle panel of Fig. 2(b)), clearly demonstrating the quasi-periodic nature of these avalanches. Quantitative comparisons of quasi-periodic avalanche characteristics for *LR* and *MR* systems are shown in Fig. 2(b). On average, *LR* system manifests larger strain intervals and more significant stress drops as compared with *MR* system. These results demonstrate that roughness strongly enhances global stress fluctuations and is a key parameter for granular avalanches [10, 11].

To understand the microscopic origin of these distinctive avalanche behaviors, we quantify the particle dynamics for systems with different roughness. It has been found that in 3D granular system microscopic nonaffine particle displacements are strongly correlated with macroscopic avalanches [16, 19, 42]. Following a similar approach [41-43], we calculate the nonaffine displacements $\boldsymbol{u}$ of each particle during shear deformation. Specifically, for each particle located at $\boldsymbol{r}$, we calculate the particle displacement $\boldsymbol{u}_0(\boldsymbol{r}) = \boldsymbol{r}(\gamma + \delta\gamma) - \boldsymbol{r}(\gamma)$. Then, we obtain the nonaffine displacement $\boldsymbol{u}(\boldsymbol{r})$ of each particle by $\boldsymbol{u}(\boldsymbol{r}) = \boldsymbol{u}_0(\boldsymbol{r}) - \frac{1}{n}\sum_{i=1}^{n}\boldsymbol{u}_0(\boldsymbol{r}_i)$, *i.e.*, $\boldsymbol{u}_0(\boldsymbol{r})$ minus the average displacement of its $n$ contacting neighbors [44]. Note that all the following results are insensitive to the specific ways the nonaffine displacement is defined. Figure 3(a) shows a snapshot of nonaffine displacement amplitude $u$ at the critical strain $\gamma_c$ for a typical avalanche event, which is spatially heterogeneous. Figure 3(b) shows $\bar{u}$ (global mean values of $u$) as a function of strain for the three systems, which demonstrates rather similar behaviors to those of global stress fluctuations. Actually, figure 3(c) shows that the stress drop $s$ is in fact directly proportional to $\bar{u}$ and falls on a master curve for both $LR$ and $MR$ systems. This reveals that avalanches and microscopic nonaffine displacements are indeed strongly correlated and that roughness enhances avalanche through microscopic nonaffine displacement. Interestingly, the inset of Fig. 3(b) shows that the nonaffine displacement fluctuations disappear if a larger strain interval (about $10\delta\gamma = 9.1\%$) is used to calculate $u$, suggesting that an avalanche is a fluctuation of nonaffine displacement and happens at a certain characteristic strain interval, which could be related to the size of surface roughness.

As shown in Fig. 2(b), the quasi-periodic avalanches display different distributions of stress drop magnitude $s$ for the three systems. To understand the physical origin of these distributions and the differences among the three systems, we calculate the 3D spatial auto-correlation function of nonaffine displacement,

$$C(r) = \langle [u(\boldsymbol{R}) - \bar{u}][u(\boldsymbol{R}+\boldsymbol{r}) - \bar{u}] \rangle_R / \langle [u(\boldsymbol{R}) - \bar{u}][u(\boldsymbol{R}) - \bar{u}] \rangle_R ,$$

where $\langle \cdot \rangle_R$ denotes spatial averaging over all particles after excluding boundary particles. As shown in Fig. 3(d), $C(\boldsymbol{r})$ is strongly anisotropic with the largest correlation along the shear direction. The corresponding angularly averaged correlation functions $C(r)$, with $r = |\boldsymbol{r}|$, for the three systems are also calculated. In general, $C(r)$ for different shear steps can be well fitted by a stretched exponential function, $\exp[-(r/\xi)^{0.5}]$, where $\xi$ is the correlation length, as shown in Fig. 3(e). The inset of Fig. 3(e) shows the pdf of $\xi$ at different shear steps for the three systems, in which a growth of $\xi$ with increasing roughness can be observed.

Furthermore, we find a linear relationship between $\xi$ and quasi-periodic avalanche stress drop $s$ for both *MR* and *LR* systems (Fig. 3(f)). This demonstrates that quasi-periodic avalanches are directly associated with spatially larger nonaffine displacement correlation, indicating the formation of active nonaffine clusters in space. This scenario is qualitatively different from the picture in which only the magnitude of nonaffine displacement for individual particle is enhanced (see Fig. S1 in the Supplemental Materials (SM) [45] for more discussion). A natural explanation for this linear relationship is that roughness will enhance geometric locking on frictional contacts which leads to a stronger inter-particle coupling [11, 46], and subsequently particle dynamics is correlated on a larger length scale which results in a stronger stress drop in an avalanche. It is natural to assume that particle shape, cohesive force, elastic interaction can also play similar roles in enhancing the interactions between particles, and therefore induce large avalanches. We also note that for the *MR* system, the correlation between $\xi$ and $s$ is poor when $s$ is small, as shown in Fig. 3(f). We speculate that this is due to the fact that crackling-like small avalanches have different physical mechanisms from the larger, quasi-periodic avalanches [26, 27, 47].

Figure 3 demonstrates that the correlation length $\xi$ is a useful parameter to characterize quasi-periodic avalanches. Therefore, to further understand the incubation pathway of a quasi-periodic avalanche and its associated phase transition characteristics, we analyze the evolution of $\xi$ as an imminent quasi-periodic avalanche is approached. In the following we mainly focus on the *LR* system, while the results for the *MR* system are qualitative similar. Specifically, for each quasi-periodic avalanche event, the correlation length $\xi$ is calculated as a function of $|\gamma - \gamma_c|$ ranging from the initiation strain $\gamma_i$, which starts at the end of the last quasi-periodic avalanche event, to the critical strain $\gamma_c$ at which the avalanche occurs. Fig. 4(b) shows the mean value of $\xi$, averaged over all quasi-periodic avalanche events, as $\gamma_c$ is approached. It shows that $\xi$ first increases slowly before abruptly jumping to a large value when the avalanche occurs at $\gamma_c$. Recently, active nonaffine clusters have been used to study avalanches in 3D granular system [42]. Thus, to support the conclusion drawn from Fig. 4(b), we also use the active nonaffine cluster size as an alternative length scale to characteristic the growth of avalanche behavior in the system. Active particles (top 20% $u$ particles) that are in contact with each other are defined to belong to the same cluster, as shown in Fig. 4(a), and the correlation length is defined by the mean cluster size $\overline{N_c}$ [42]. Figure 4(c) shows the evolution of $\overline{N_c}$ as a function of $|\gamma - \gamma_c|$, and it shows a similar trend as $\xi$ in Fig. 4(b). We have checked that this result is robust when other proportions, *i.e.*, top 25%, 30%, 40% of active particles are used (see Fig. S2 in SM [45]). The similar behavior of $\xi$ or $\overline{N_c}$, *i.e.*, a very slow evolution until an abrupt increases at critical point, suggests a spinodal nucleation picture of a first-order phase transition for the quasi-periodic avalanche [29, 37, 48] .

As originally concluded from rotating drum experiments, a granular pile is stable if the slope angle is less than the repose angle $\theta_r$ [28]. Beyond $\theta_r$, the system is metastable until the maximum instability angle $\theta_s$ is exceeded, at which a system-spanning granular avalanche occurs. Physically, the increasing slope of the sandpile leads to the tilting of the potential energy landscape in which the system sits in a potential well with a finite energy barrier. When the slope is further increased to $\theta_s$, the energy barrier disappears giving rise to a global instability and as a result an avalanche occurs [28]. Here $\theta_s$ plays the similar role of "spinodal point" of a first-order phase transition [29, 47, 48]. In our study, a similar physical scenario can be invoked. As shown in Fig. 4, the slow evolution of $\xi$ or mean cluster size $\overline{N_c}$ before the avalanche is triggered resembles heterogeneous nucleation in the metastable phase, and the abrupt increases of $\xi$ near the critical strain $\gamma_c$ is analogous to a spinodal instability as the shear strain pushes the system over the potential well of the metastable states [29, 37, 48].

We note that the active nonaffine clusters are heterogeneously distributed in space suggesting the possible formation of shear bands, as shown in Fig. 4(a). Consequently, the nucleation of avalanche mainly happens in the shear bands. However, all three systems display similar shear band structures as shown in Fig. S3 of SM [45], and therefore, shear bands cannot be the cause for the different avalanche behaviors. Instead, we speculate that in our system avalanches result from force chain reorganization dynamics [15, 49], which happens on the length scale of surface roughness [50, 51]. This by nature is different from the particle-scale dynamics which involves significant particle rearrangement to form the shear band. This view is supported by the fact that the nonaffine displacement of individual particle during an avalanche is very small (only 1~5% *D*, see Fig. 3(b) or Fig. S1 in SM [45]), although their spatial correlation extends to a length scale of several particles. This also naturally explains why the volume fraction $\phi$ shows a smooth dependence on strain, see Fig. 2(a). Physically, both particle and surface roughness length scales exist in granular materials and the latter is directly related to contact and force structures [49, 52]. When scale separation is large, mechanical fluctuations induced by force chain reorganization is weak and can in principle be properly coarse-grained to establish a constitutive relationship for granular materials based only on structural information on the particle level [43, 53, 54]. In the other case, when weak scale separation occurs, *e.g.*, large surface roughness, the mechanical fluctuations can be collectively excited, which results in an avalanche behavior visible on the particle length scale as seen in the current work.

In summary, we investigate how the roughness of the particles influences the properties of avalanches in a sheared system. We find that systems with rough particles display large avalanches while such events are almost absent in systems of smooth particles. Despite the fact that the *MR* system displays qualitatively similar avalanche behaviors as the *LR* system, the avalanche strain interval and stress drop size are quantitatively very different as shown in Fig. 2(b). This is most likely due to the different degree of roughness of two systems, which determines the energy relaxation length scale. Therefore, it is reasonable to speculate that avalanches also happen in systems with small roughness, *e.g.*, the *SR* system, in which avalanche interval and stress drops are too small to be experimentally visible. Finally, it is well-known that quasi-periodic avalanches or stick-slip behaviors are generic phenomena in earthquake faults, so that our result can shed some light on the nature of quasi-periodic earthquakes [6, 24, 27].

The work is supported by the National Natural Science Foundation of China (No. 11974240) and Shanghai Science and Technology Committee (No. 19XD1402100).

\* Corresponding author

yujiewang@sjtu.edu.cn

cjxia@phy.ecun.edu.cn

different strains, mean cluster size $\overline{N_c}$ vs $|\gamma-\gamma_c|$ for different definitions of active nonaffine clusters, shear band behavior, the effective friction coefficients of the three systems, and estimating the average strain interval based on the length scale of roughness.

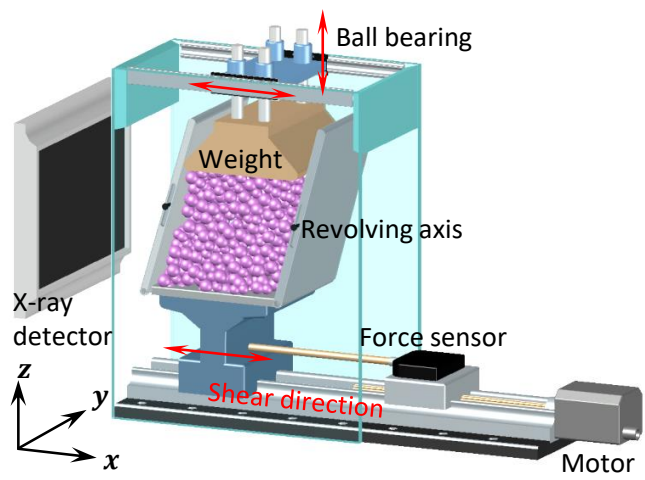

FIG. 1 Schematic of the experimental setup.

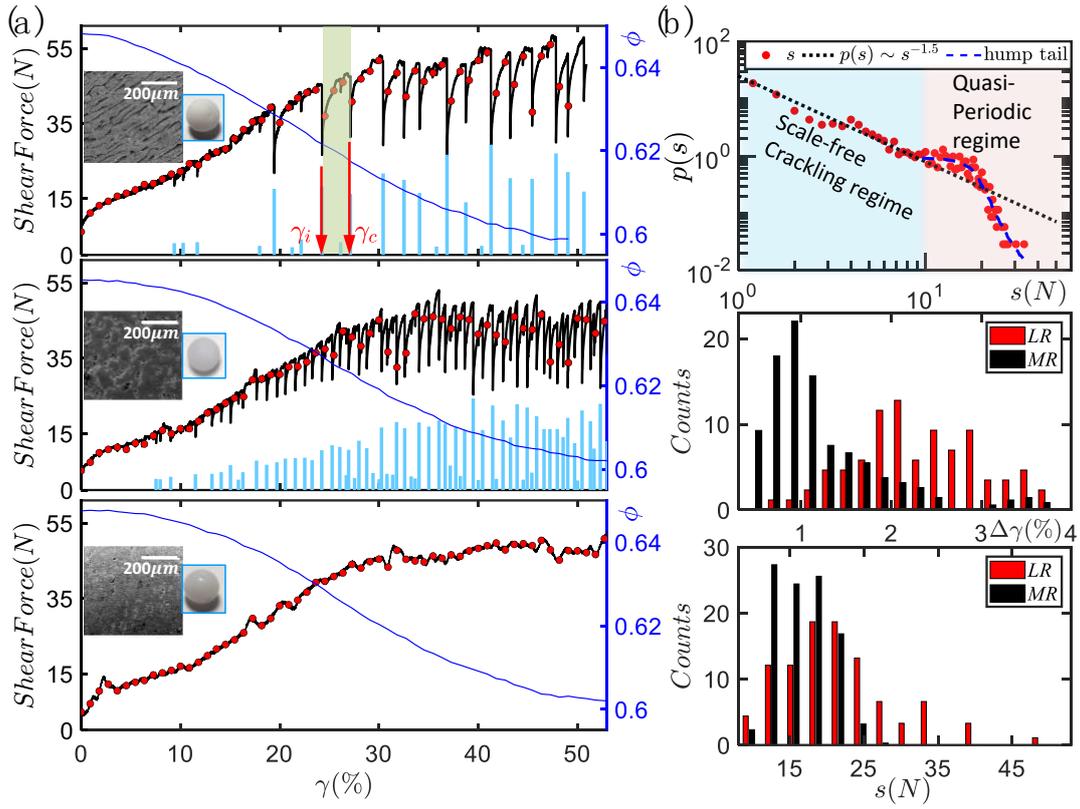

FIG. 2 (a) Global shear force (black curve) and volume fraction $\phi$ (blue curve) as a function of strain $\gamma$ for the three systems (from top to bottom: *LR, MR, SR*). The red circles mark strains where CT scans are taken. The height of blue bar labels the shear force drop $s$. Insets, the camera (and SEM) images of the particle (surface). (b) Top panel: pdf of $s$ for the *MR* system obtained over one thousand avalanche events. The black dotted line shows a power law of $s^{-1.5}$ and the blue-dashed line guides the hump tail. Middle and bottom panels: the histograms of strain interval $\Delta \gamma$ and $s$ for avalanches within the hump tail.

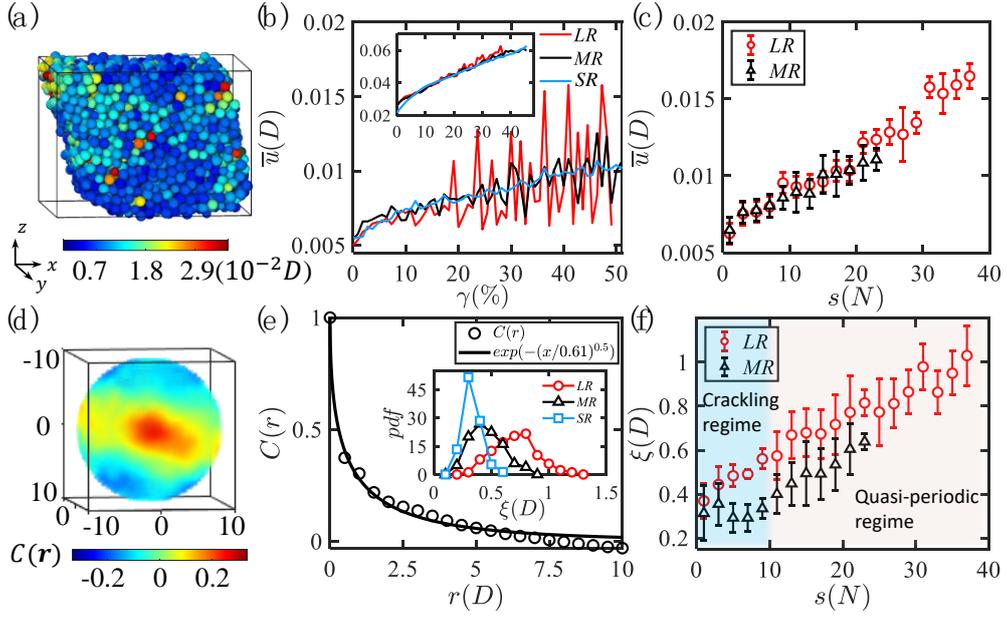

FIG. 3 (a) Snapshot of nonaffine displacement amplitude $u$ at a typical $\gamma_c$. (b) Mean nonaffine displacement $\bar{u}$ as a function of strain $\gamma$. Inset, $\bar{u}$ calculated by ten times larger strain step interval ($10\delta\gamma$). (c) $\bar{u}$ vs stress drop $s$. (d) The spatial auto-correlation map $C(\mathbf{r})$ of (a). (e) Angular averaged correlation function $C(r)$ of (d), and the corresponding stretched exponential fit of $\exp[-(x/\xi)^{0.5}]$. Inset, the pdfs of $\xi$ for *LR, MR, SR* systems. (f) Correlation length $\xi$ vs *s*.

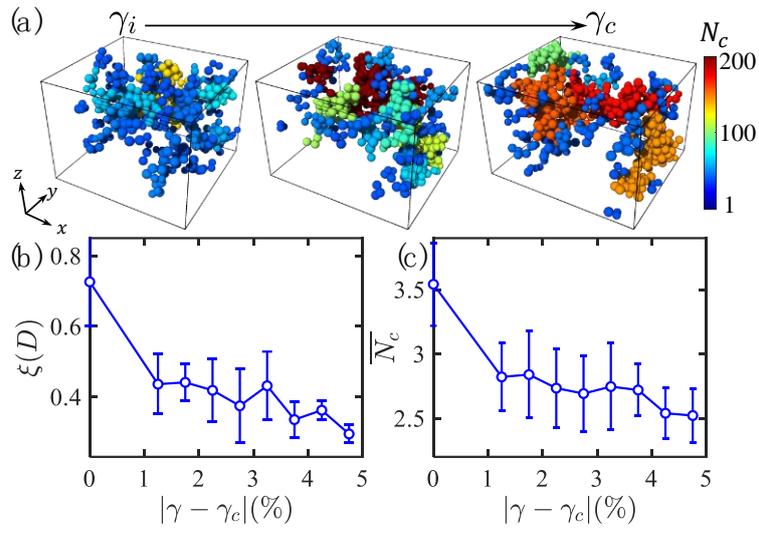

FIG. 4 (a) Snapshots of active nonaffine clusters (top 20% of $u$) at different strains of a typical quasi-periodic avalanche event. Particles in the same cluster are labeled by the same color, and for the sake of clarity, clusters smaller than three are not shown. (b) The correlation length $\xi$ as a function of $|\gamma - \gamma_c|$. (c) The average cluster size $\overline{N_c}$ as a function of $|\gamma - \gamma_c|$.

# Supplemental Material for
# *Influence of Roughness on Granular Avalanches*

Jie Zheng[1], Yi Xing[1], Ye Yuan[1], Zhifeng Li[1], Zhikun Zeng[1], Shuyang Zhang[1], Houfei Yuan[1], Hua Tong[1], Chengjie Xia[2*], Walter Kob[1,3], Xiaoping Jia[4], Jie Zhang[1,5] and Yujie Wang[1,6*]

[1]*School of Physics and Astronomy, Shanghai Jiao Tong University, Shanghai 200240, China*
[2]*School of Physics, East China Normal University, Shanghai 200241, China*
[3]*Laboratoire Charles Coulomb, University of Montpellier and CNRS, Montpellier 34095, France*
[4]*Institut Langevin, ESPCI Paris, Université PSL, CNRS, Paris 75005, France*
[5]*Institute of Natural Sciences, Shanghai Jiao Tong University, Shanghai 200240, China*
[6]*Materials Genome Initiative Center, Shanghai Jiao Tong University, Shanghai 200240, China*

\* Corresponding author

yujiewang@sjtu.edu.cn

cjxia@phy.ecun.edu.cn

1. **Pdfs of particle nonaffine displacements at different strains**

The pdfs of nonaffine displacement amplitude $u$ and its rescaled counterpart $u/\overline{u}$ at different strains of a typical avalanche event are shown in Fig. S1. Here, $\overline{u}$ is the global mean value of $u$. Figure S1(a) shows that $u$ is about one percent of the particle diameter, thus much smaller than the scale of the particles. Hence this demonstrates that avalanches (stress release) mainly happen on the length scale of the contact-level roughness. The pdf of $u/\overline{u}$ in Fig. S2(b) do not collapse onto a same master curve before nor when an avalanche occurs, indicating that avalanche does not happen via a simple overall enhancement of $u$ for all particles.

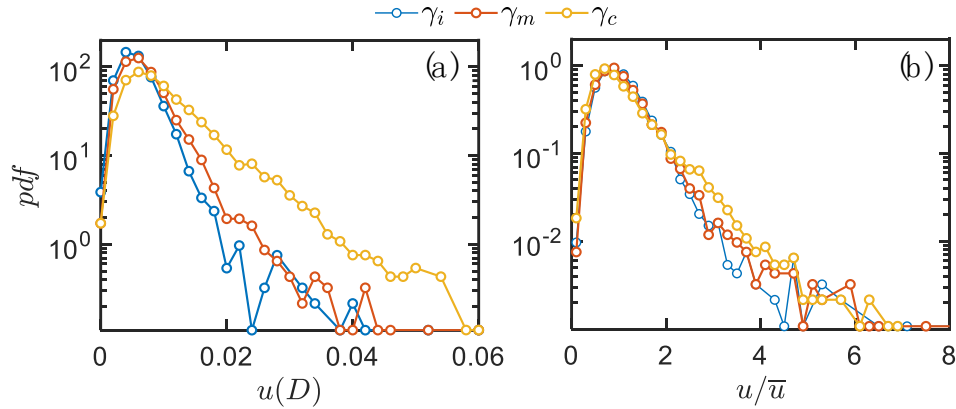

FIG. S1 (a) Pdfs of nonaffine displacement magnitude $u$; (b) Pdfs of rescaled $u/\overline{u}$. With $\gamma_i$, $\gamma_m$, $\gamma_c$ denote the initial, medium ($\gamma_m \approx (\gamma_c - \gamma_i)/2$), critical strain of a typical avalanche, respectively. The unit $D$ is the mean particle diameter.

## 2. Mean cluster size $\overline{N_c}$ vs $|\gamma - \gamma_c|$ for different definitions of active nonaffine clusters

In Fig. 4(c) of the main text we have shown that the mean cluster size $\overline{N_c}$ increases quickly when the strain approaches the critical value $\gamma_c$. In order to show that this behavior is not influenced by the exact definition of $\overline{N_c}$, we show this size in Fig. S2 by using the particles which have nonaffine displacements that are in the top 20%, 30%, and 40% of their distribution. For all these cases, the curves of $\overline{N_c}$ vs $|\gamma - \gamma_c|$ show a qualitative similar behavior: first a slow, and then an abrupt rise to a large value at the critical strain $\gamma = \gamma_c$ when avalanches occur. These results indicate that the spinodal nucleation picture claimed in main text is robust.

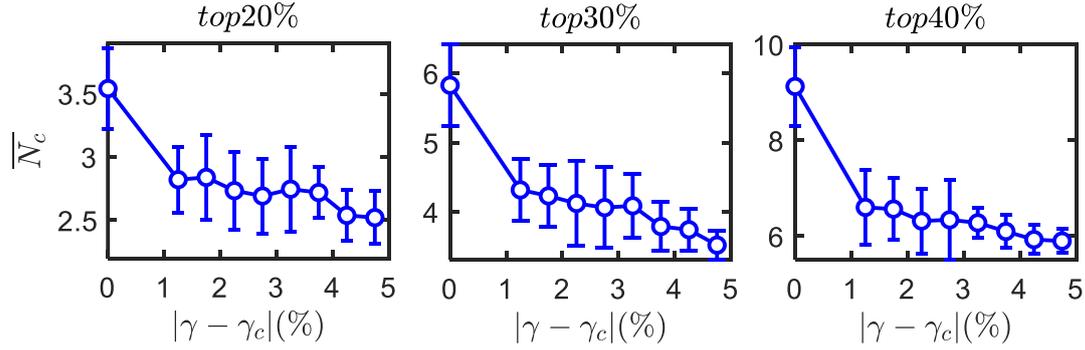

FIG. S2 From left to right, the mean cluster sizes $\overline{N_c}$ vs $|\gamma - \gamma_c|$ for active nonaffine clusters that consist of top 20%, 30% and 40% nonaffine displacement particles.

## 3. Shear band behavior

In Fig. S3 we show the spatial distributions of active nonaffine clusters for the three systems with different roughness. The spatial features are qualitatively similar: the clusters are distributed heterogeneously in space and have fractal-like structure. The cluster aggregated regime indicates the existence of shear bands. The fact that the shear bands are very similar and independent of the roughness indicates that shear bands are not a relevant factor in determining avalanches in our system.

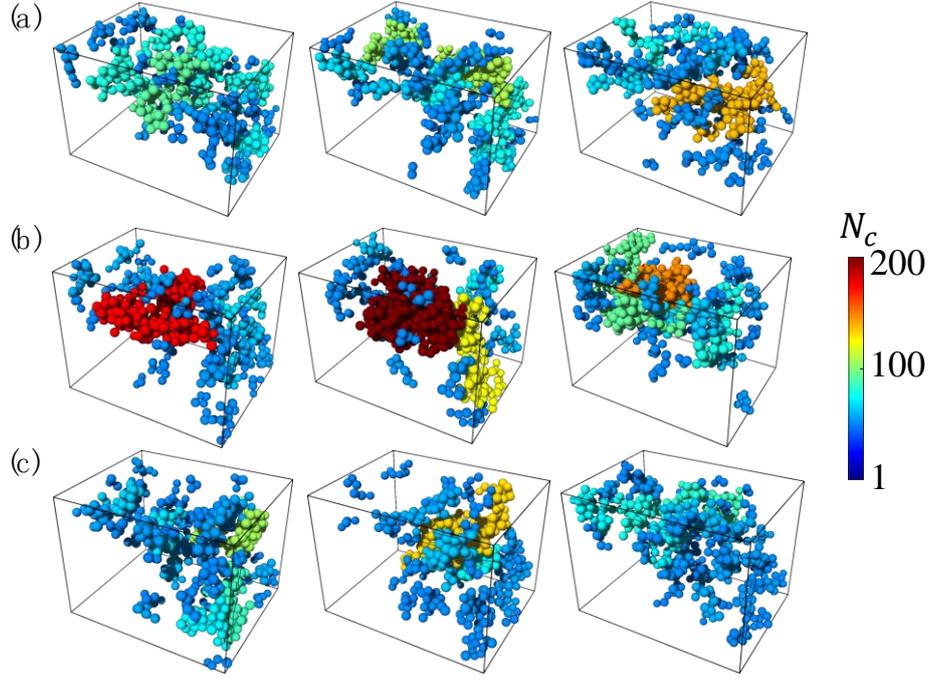

FIG. S3 Snapshots of active nonaffine clusters (consisting of top 20% nonaffine displacement particles) for the three systems (from (a) to (c): *LR*, *MR*, *SR*). For each system, three sub-panels are arbitrary selected from steady states with $\gamma$=40%, 44.5%, 49.1%, respectively. The particles in the same cluster are labeled by the same color, and for the sake of clarify clusters with cluster size $N_c$ less than three are not shown.

## 4. The effective friction coefficients of the three systems

The global stress-strain curves in Fig. 2(a) show that the global stresses (shear force) at steady states of the three systems appear to be quite similar despite their distinctive avalanche behaviors. This indicates similar effective friction coefficients for all three systems. To further validate this observation, we use a rotating drum to measure the repose angle $\theta_r$ of granular packing and evaluate the effective friction coefficient $\mu$ of the three systems as $\mu$ can be calculated by $\mu = tan(\theta_r)$. The rotating drum contains nearly 6000 particles, which is slowly rotated until an avalanche occurs and the resulting $\theta_r$ is measured by a laser protractor. For each type of particle, we did 100 measurements. The mean friction coefficients obtained are 0.68, 0.65, 0.61, for *LR, MR, SR* system, respectively. However, the similar effective friction coefficients suggest that in our system avalanches are mainly controlled by the surface roughness rather than the friction itself. This thus cautions to discriminate these two terms since in the past they have been often used interchangeably [46].

**5. Estimating the average strain interval based on the length scale of roughness**

In a previous study, the length scale of asperity on the particle surface was used to estimate the activation energy of stick-slip behavior [50]. Following a similar approach, we can evaluate the strain interval (mean avalanche duration) $\Delta\gamma$ of the quasi-periodic avalanche. In the simplest approximation, $\Delta\gamma$ should be related to the strain required to overcome an asperity between contacting particles since avalanches appear to a collective process of unitary asperity overcoming events, *i.e.*,

$$\Delta\gamma \sim \frac{h}{D/2}. \tag{1}$$

Here, $D$ is the mean particle diameter and $h$ is the average size of roughness asperity. Equation (1) suggests that the mean strain interval $\Delta\gamma$ of avalanche is directly related to the length scale of surface roughness $h$. On average, *LR* system should display larger strain interval as compared with *MR* system, which is self-consistent with the visual inspection of the roughness size result in middle panel of Fig. 2(b) of the main text.